\documentclass[final,5p,times,twocolumn,authoryear]{elsarticle}

\usepackage{amssymb}
\usepackage{lipsum}
\usepackage{graphicx}
\usepackage{hyperref}
\usepackage{amsmath}
\usepackage{color}
\usepackage[caption=false]{subfig}
\usepackage{comment}
\usepackage{verbatim}
\usepackage{tabularx}
\usepackage{booktabs}
\usepackage{multirow}
\usepackage{ulem}

\journal{New Astronomy}

\newcommand{\pine}{{\sc PineForest}}
\newcommand{\iso}{{\sc Isolation Forest}}

\newcommand{\aad}{{\sc Active Anomaly Discovery}}

\title{Dataset of artefacts for machine learning applications in astronomy}

\begin{document}


\author[first]{Sreevarsha Sreejith}
\author[second,third]{Maria V.~Pruzhinskaya}
\author[fourth]{Alina A.~Volnova}
\author[fifth]{Vadim V.~Krushinsky}
\author[sixth]{Konstantin L.~Malanchev}
\author[third]{Emille E.~O.~Ishida}
\author[second]{Anastasia D.~Lavrukhina}
\author[second,seventh]{Timofey A.~Semenikhin}
\author[third]{Emmanuel Gangler}
\author[second,eighth]{Matwey V.~Kornilov}
\author[ninth]{Vladimir S.~Korolev}

\affiliation[first]{organization={Physics Department, University of Surrey},
addressline={Stag Hill Campus}, 
city={Guildford},
postcode={GU2 7XH}, 
state={Surrey},
country={UK}}

\affiliation[second]{organization={Lomonosov Moscow State University, Sternberg Astronomical Institute},
addressline = { Universitetsky pr.~13},
city = {Moscow},
postcode={119234},
country={Russia}}

\affiliation[third]{organization={Universite Clermont Auvergne},
addressline = {CNRS, LPCA},
city = {Clermont-Ferrand},
postcode={F-63000},
country={France}}

\affiliation[fourth]{organization={Space Research Institute of the Russian Academy of Sciences (IKI)},
addressline = {84/32 Profsoyuznaya Street},
city = {Moscow},
postcode={117997},
country={Russia}}

\affiliation[fifth]{organization={Laboratory of Astrochemical Research, Ural Federal University},
addressline = {Ekaterinburg, ul. Mira d. 19},
city = {Yekaterinburg},
postcode={620002},
country={Russia}}

\affiliation[sixth]{organization={McWilliams Center for Cosmology \& Astrophysics, Department of Physics},
addressline = {Carnegie Mellon University},
city = {Philadelphia},
postcode={PA 15213},
country={USA}}

\affiliation[seventh]{organization={Lomonosov Moscow State University, Faculty of Physics},
addressline = {Leninskie Gory 1-2},
city = {Moscow},
postcode={119991},
country={Russia}}

\affiliation[eighth]{organization={National Research University Higher School of Economics},
addressline = {21/4 Staraya Basmannaya Ulitsa},
city = {Moscow},
postcode={105066},
country={Russia}}

\affiliation[ninth]{Independent researcher}

\begin{abstract}

Accurate photometry in astronomical surveys is challenged by image artefacts, which affect measurements and degrade data quality.
Due to the large amount of available data, this task is increasingly handled using machine learning algorithms, which often require a labelled training set to learn data patterns. We present an expert-labelled dataset of 1127 artefacts with 1213 labels from 26 fields in ZTF DR3, along with a complementary set of nominal objects. The artefact dataset was compiled using the active anomaly detection algorithm \pine,\ developed by the SNAD team. These datasets can serve as valuable resources for real-bogus classification, catalogue cleaning, anomaly detection, and educational purposes. Both artefacts and nominal images are provided in FITS format in two sizes ($28 \times 28$ and $63 \times 63$ pixels). The datasets are publicly available for further scientific applications.

\end{abstract}



\maketitle
\section{Introduction}

With the  advent of existing wide-field surveys like the Zwicky Transient Facility (ZTF, \citealt{2019PASP..131a8002B}),  the Euclid Mission \citep{euclid2010} and the James Webb Space Telescope \citep{jwst2006} and upcoming surveys such as the Vera C. Rubin Observatory Legacy Survey of Space and Time (LSST, \citealt{lsst2019}), 
preparing  astronomical data for human analysis has become an increasingly challenging task. Since the amount of data makes 
individual object inspection prohibitive, this is achieved through efficient algorithms and high performance computing, often incorporating machine learning techniques. Most surveys have an initial filter that removes objects or features that are deemed spurious in images and catalogs as part of their data processing pipeline (e.g.,~\citealt{2012ExA....33..173K,duev2019}). 
However, this step is not entirely foolproof, leaving artefacts such as reflection ghosts, bad columns, satellite tracks, as well as photometric contamination of real objects due to light from nearby bright stars or closely passing asteroids \citep[e.g.][]{malanchev2021, 2021AJ....162..206S,Pruzhinskaya2023}.
These artefacts can appear in some exposures and disappear in others, affecting astrometric calibration and the photometry of nearby objects, ultimately degrading data quality. Thus, their identification and removal are of paramount importance before the compilation of science-ready catalogs.

There has been some limited effort to detect and/or mask artefacts in images, focusing on data from different surveys such as \cite{desai2016}, \cite{chang2021} and \cite{tano2022}. However, most of the related work uses statistical/deep learning  algorithms to classify objects into `real' or `bogus' such as \citealt{duev2019,paranjpye2019,alerce2021,killestein2021,Semenikhin2024,weston2024}. 
A common requirement for this approach is a labeled dataset for classifier training, which is not always easy to obtain. While image artefacts can be simulated (e.g., \citealt{2023A&A...679A.135P}), visual inspection of real data remains the most direct way to obtain accurate labels. However, labeling is a time-consuming and resource-intensive process. In recent years, this task has been partially delegated to citizen science initiatives such as Zooniverse\footnote{\url{https://www.zooniverse.org/}} (formerly Galaxy Zoo), its transient-specific branch Zwickyverse\footnote{\url{https://www.zooniverse.org/projects/rswcit/zwickys-quirky-transients}}, the Jupiter-specific project Jovian Vortex Hunter \citep{JVH}, and others. 
While citizen science projects can produce large amounts of labelled data, this will generally be accompanied by large uncertainties requiring careful statistical modelling before their results can be used in subsequent studies. Labels provided by astronomers -- though fewer -- are likely to be more accurate and thus better suited as input for machine learning algorithms. Moreover, experts can identify the most probable cause for a given artefact, thus also contributing to the diagnostic and improvement of instrument operations. Even a relatively small data set, compiled by experts, can be a valuable starting point for machine learning algorithms. For example, \citealt{Semenikhin2024} demonstrated that high real-bogus classification performance can be achieved starting from just over 3000 expert-labelled examples for training.
Thus, the goal of this work is to provide expert-labeled datasets of both artefacts and non-artefacts (hereafter referred to as nominals) from the ZTF survey. These datasets can be used for scientific tasks such as real-bogus classification and anomaly detection, as well as for education and outreach. 

Since 2018, the SNAD\footnote{\url{https://snad.space}}\citep{volnova2024} team has been working on different aspects of anomaly detection in astrophysical data, leading to the development of the \texttt{coniferest}\footnote{\url{https://coniferest.snad.space/}} package \citep{coniferest}, which incorporates several anomaly detection algorithms, including \iso \ \citep{isoforest}, \aad \ \citep{das2017,aadpaper}, and \pine\ (Korolev et al., in prep.). These algorithms have been applied to studies of transients and variable objects, including supernovae, variable stars, and red dwarf flares \citep{pruzhinskaya2019,malanchev2021,Pruzhinskaya2023,VL2024}. The SNAD team's efforts have also resulted in the creation of the SNAD anomaly knowledge base \citep{snadviewer}, which provides labels for thousands of ZTF objects, which, in turn, has been the motivation for this work.  

This paper is organised as follows. Section \ref{sec:data} describes the data selection process. Section \ref{sec:arts} defines the labels used in this work according to the SNAD labelling schema. Section \ref{sec:method} details the algorithm and its implementation.  Sections \ref{sec:results}-\ref{sec:disc} present the results and their discussion, respectively. Section \ref{sec:summconc} provides a summary and conclusions.

\section{Data} 
\label{sec:data}

The Zwicky Transient Facility is a wide-field survey that scans the northern sky to detect, characterize, and catalog photometric features of optical transients in the \textit{g}, \textit{r}, and \textit{i} bands (referred to as \textit{zg}, \textit{zr}, and \textit{zi}) using a custom-built mosaic camera mounted on the Samuel Oschin Telescope at Palomar Observatory and has been operational since 2018~\citep{2019PASP..131a8002B}.

The data used in this work is part of the third ZTF data release (DR3), which includes approximately 9.4 months of observations from March to December 2018 \citep{ZTFDR3}. \textit{zr}-band light curve data for $182$ fields were selected from this release based on two quality cuts: (1) the object has at least $100$ observations in the \textit{zr}-band, and (2) the object quality is good according to ZTF pipeline\footnote{\texttt{catflags=0} in ZTF DRs.}. ZTF object identifiers (OIDs), which are unique to each band and filter, are used to identify objects within these fields.

\begin{figure*}
    \centering
    \includegraphics[width=\textwidth]{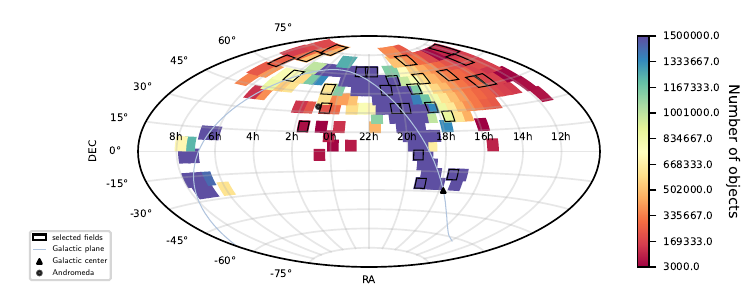}

   \caption{Distribution of the 182 feature-extracted fields from ZTF DR3 selected by applying the quaity cuts mentioned in Section \ref{sec:data}, color-coded by the number of objects in each. The black rectangles indicate the fields selected for this study.}
    \label{fig:fields}
\end{figure*}

From these 182 fields, we selected a representative subset of 26 fields covering the ZTF footprint, as shown in Figure~\ref{fig:fields}. To ensure a balanced selection, we first sorted the fields by the number of objects they contain and divided them into batches of seven. From each of the 26 batches, one field was randomly chosen. These selected fields were then analysed using the \pine~anomaly detection algorithm (Section~\ref{sec:PF}). The number of objects per field varies from $38,623$ to $9,486,393$.

The light curve data for selected fields was processed using the \texttt{light-curve}\footnote{\url{https://github.com/light-curve/light-curve-python}} package, a time-series feature extraction tool developed as part of the SNAD pipeline (\citealt{malanchev2021,lcpaper}). The extracted feature set consists of the same 42 \textit{zr}-band features used in \citealt{Pruzhinskaya2023}.

\section{Artefact labelling system} 
\label{sec:arts}

In \citealt{malanchev2021}, 68\% of the detected outliers were found to be bogus light curves that arise from detector/optical system, image subtraction issues, blending, and other effects. In this work, we group these bogus detections under the umbrella of `artefacts' and try to isolate them in the feature-extracted light curve dataset.

To systematically label objects detected during our anomaly detection efforts, SNAD has developed its own tagging schema, which we follow in this work. The detailed definitions of the tags used in this dataset are provided in Section \ref{sec:defns}. This labeling system is self-consistent, designed to address SNAD specific challenges and requests, and may not fully align with classification schemes commonly used in the broader astronomical community. 

We label only those artefacts that directly affect the central object and its photometry. Although different image defects can  introduce additional artefacts within the cutout, we consider this approximation a reasonable starting point. Notably, the appearance of artefacts can vary depending on the cutout size -- an artefact visible in a larger image may be absent in a smaller one. Introducing additional labels for non-central artefacts could lead to inconsistencies and misclassifications in the SNAD anomaly knowledge base. In order to maintain consistency and relevance across different tasks, we choose to label only the central object.

\subsection{Types of artefacts} 
\label{sec:defns} 

\begin{figure*}[!htbp]    
    \centering
    \includegraphics[width=\textwidth]{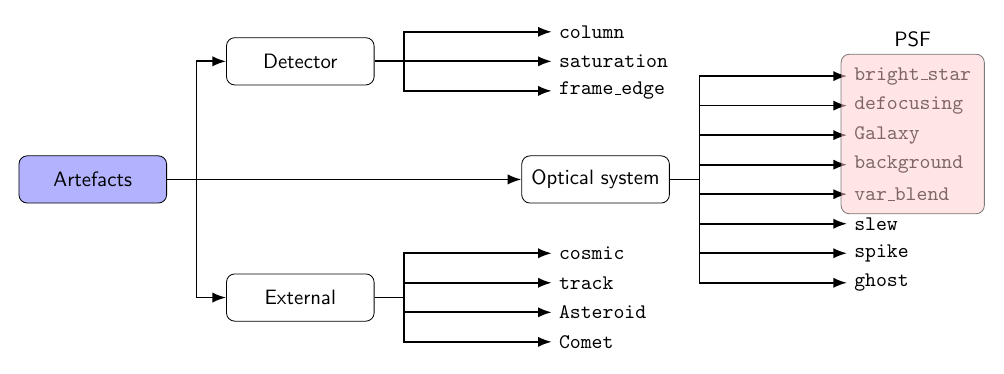}
   \caption{Figure showing the artefact labelling schema adopted in SNAD, as described in Section \ref{sec:arts}. All selected objects are tagged by the general label `artefact' and then by the sub-labels  according to their characteristics.}
    \label{fig:schema}
\end{figure*}

In the SNAD anomaly knowledge base, artefacts are categorized based on their origin, a requirement put in place to address the demands of the project and its collaborators. These form three broad groups of artefacts from the ZTF data: those originating from the detector, from the optical system, and those resulting from external factors (see Fig.~\ref{fig:schema}).

\subsubsection{Detector artefacts}

This category includes artefacts that arise from imperfections in the CCD detector. It contains the following tags:
\begin{itemize}[-]
    \item \texttt{column} describes bad pixel columns in the CCD that malfunction primarily due to production process issues. For example, some pixels do not turn on at all and stay dark/dead, while others always return hot/maximum signal. We also include traps under this category. Traps are damaged pixels that do not transfer the contained electrons properly when they are read out, causing partial data loss along the column.
    \item \texttt{saturation} occurs when a pixel exceeds its electron-holding capacity, leading to charge overflow into the neighboring pixels. It appears as a white streak or tail in images, and mostly occurs in very bright stars. \item \texttt{frame\_edge} describes artefacts where a targeted object is located at the edge of the detector field of view.
\end{itemize}

\subsubsection{Optical system artefacts}

These artefacts are caused by the telescope's optical system and are identified by the tags:
\begin{itemize}[-]
        \item \texttt{bright\_star} refers to the proximity of an 
        object whose scattered light contaminates the target and cannot be accurately subtracted.
        \item \texttt{defocusing} results from overlapping PSFs of nearby stars leading to blending or from optical defocusing. 
        \item \texttt{Galaxy} denotes a galaxy being the primary target. In this case, PSF photometry is wrong since galaxies are not point sources. Moreover, most artefacts of this type also exhibit slight defocusing, further altering the PSF.   
        \item \texttt{background} refers to irregular background gradients caused by improper sky subtraction.
        \item \texttt{var\_blend} occurs when scattered-light echoes from nearby variable stars affect light curves of the primary target, introducing variability (see Section~\ref{mira_echoes}).  
        The origin of this effect is the same as the \texttt{bright\_star} artefact.
        \item \texttt{slew} refers to the patterns created as the telescope slews, i.e. when the mount is moved to point to a particular position in the sky. They often appear as long tails to bright stars or as criss-cross patterns in the image.
        \item \texttt{spike} refers to diffraction spikes caused by light bending around the support struts of the secondary mirror (spider), often seen around bright stars.
        \item \texttt{ghost} is a term used for internal reflection within lenses, filters and on its barrels. It comes in various shapes and sizes, the most common of which in our data is as a smudge around the targeted object. 
\end{itemize}

\subsubsection{External artefacts}

External artefacts originate from sources outside the detector and optical system. The corresponding tags are:
\begin{itemize}[-]
    \item \texttt{cosmic} refers to cosmic rays striking the detector, which create single bright pixels or chains of bright pixels depending on the strike angle.
    \item \texttt{track} describes artefacts caused by aircraft or satellites crossing the field of view.
    \item \texttt{Asteroid} and \texttt{Comet} describe the cases when these celestial bodies overlap with the primary target, affecting its photometry.
\end{itemize}

\section{Method} 
\label{sec:method}

In this section, we describe the algorithms and methodologies used to create both the artefact and nominal datasets presented in this work.

\subsection{\pine}
\label{sec:PF}
\pine \ is an approach to active anomaly detection (AD)  developed by the SNAD team,  which has the widely used \iso \ \citep{isoforest} algorithm as a foundation. \iso \ is an AD spin on the decision tree method (\citealt{breiman2001,loh2014,hunt1966}) which detects outliers under the assumption that anomalies are frequently isolated from the bulk of the data in large data sets. 
It is also the base algorithm in feedback-based (active) AD software currently in use in astronomy such as \texttt{coniferest}~(\citealt{coniferest}; Korolev et. al. in prep.) and \texttt{ASTRONOMALY} \citep{Lochner2021}.

\pine\  
was inspired by the \aad \ algorithm of \citet{das2017}. However instead of altering the basic structure of the trees, it discards altogether trees that quickly isolate objects which do not agree with the expert definition of anomaly.  
When a session is started in \pine, an initial \iso  \ is constructed using the input data. The user is then presented with the object that is assigned the highest outlier score and is prompted to determine whether it is `anomaly' (scientifically interesting) or `regular' (already known or not interesting for this study). If the user marks the object as an `anomaly', 
the trees that assigned the status `regular' to this object are discarded, new trees are grown in their place and the next iteration occurs. 

In each step of the human-machine iteration loop, all labels collected so far are used to retrain the forest.
The retraining process occurs in three stages: new trees are built on the original unlabeled dataset; both old and new trees are evaluated based on their ability to predict labeled data; trees with poor performance are removed, ensuring that the total number of trees remains the same as in the original forest.
This iterative process of questions and filtering continue until a user-defined number of loops/budget are completed. 

\pine \ is publicly available as a part of the \texttt{coniferest} suite along with the SNAD implementation of \iso \footnote{The coniferest version of \iso \ is more efficient than the \texttt{sklearn} version and provides options for multithreading in processing. See Figure~1 and Section 4 in \citealt{coniferest} for a detailed comparison.} and the \aad \  algorithms.

\subsection{Dataset of artefacts}

As mentioned earlier, anomalies are a minority in any given dataset, making their random isolation time-consuming. This challenge motivates the use of an active anomaly detection algorithm, where prior decisions influence subsequent selections.

To that end, we run a \pine\ session with a budget of 50, selecting this many objects from each of the 26 chosen fields (1300 objects in total). Objects whose light curves exhibit atypical features corresponding to image artefacts (as defined in Section \ref{sec:arts}) are labelled as anomalies. The \pine\ parameters were set as follows: the number of trees to keep, \textit{n\_trees}, was 268, and the number of additional trees to be grown, \textit{n\_spare\_trees}, was 768, with all other parameters set to default values.

The main output of a \pine\ session is a list of ZTF OIDs, anomaly scores, and user labels. To ensure labelling consistency, the detected anomalies were further inspected by a team of four experts. This review was conducted using \texttt{SNAD ZTF Viewer}\footnote{\url{https://ztf.snad.space/}} \citep{snadviewer}, a web interface developed by the SNAD team that collates data, including light curves \& \texttt{FITS} images for ZTF DRs, from multiple public databases where different aspects of the same object can be simultaneously inspected.

The final artefact dataset includes these curated objects, which have been labelled based on their light curves and corresponding image features.

\subsection{Dataset of nominal objects}

To highlight the differences between regular objects and artefacts, we also provide a nominal dataset containing the same number of objects as artefacts, selected from the same fields. These objects were randomly chosen from each field and verified as normal through a careful inspection of their light curves and \texttt{FITS} images from ZTF. Importantly, this dataset consists of images where no artefacts are present.

\section{Results} 
\label{sec:results}

\begin{figure*}
    \centering
    \includegraphics[width=0.71\textwidth]{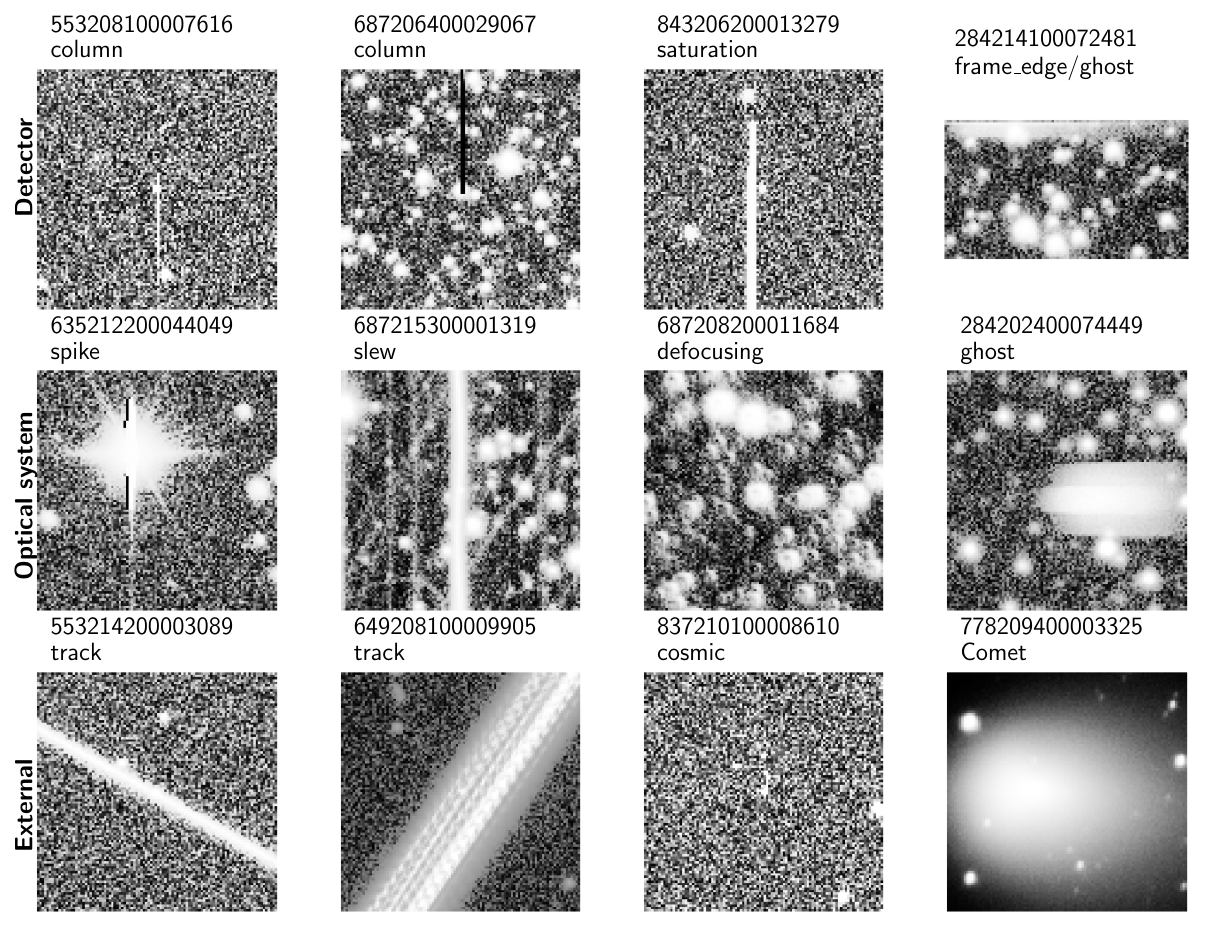}
   \caption{Figure showing examples of different types of artefacts isolated using \pine\ and labelled according to the schema defined in Section~\ref{sec:arts}. The respective ZTF OIDs are displayed above each image cutout, along with their corresponding tags. The rows represent the three main label categories. The images are $100 \times 100$ pixels each in size and the astrometric directions follow the usual convention (North up, East left).} 
    \label{fig:arts}
\end{figure*}

\begin{figure*}
    \centering
    \includegraphics[width=0.71\textwidth]{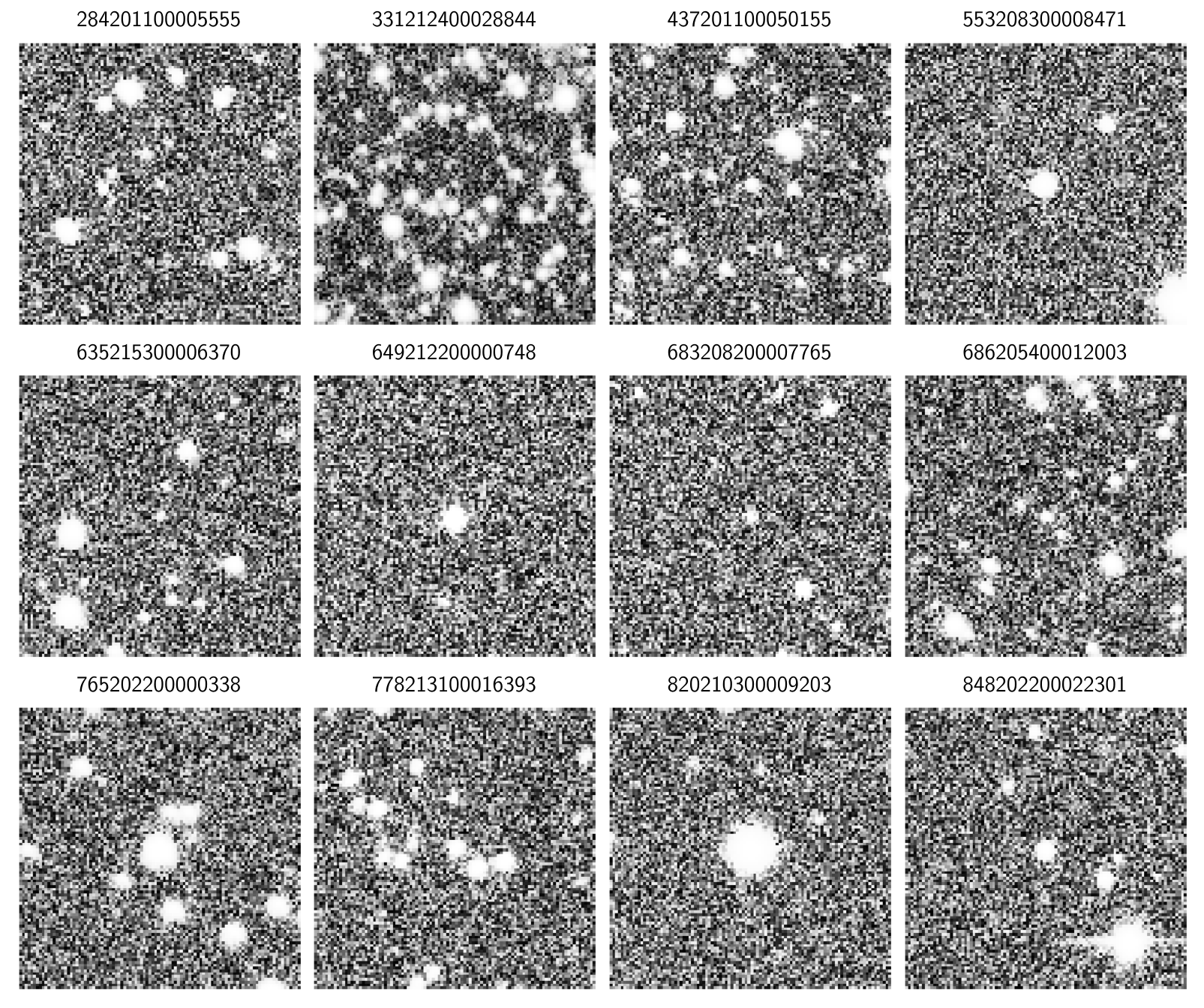}
   \caption{Figure showing some examples from the nominals dataset with their respective ZTF OIDs. The images are $100 \times 100$ pixels each in size and the astrometric directions follow the usual convention (North up, East left).}   
    \label{fig:nonarts}
\end{figure*}

Out of the 50 objects examined in each of the 26 selected ZTF fields, not all were artefacts as defined in Section \ref{sec:arts}. The exact number of images containing artefacts per ZTF field, along with the field's central coordinates, is provided in Table~\ref{tab:fields}. Examples of artefacts from different subtypes, along with their ZTF OIDs, are shown in Figure~\ref{fig:arts}. The nominal dataset contains an equal number of normal objects selected from the same fields, with some examples shown in Figure~\ref{fig:nonarts}.

\begin{table}
\scalebox{0.85}{
    \centering
    \begin{tabular}{c|c|c}
     ZTF field & Centre coordinates in deg& Number of artefacts/  \\
     & (ra, dec) &nominals \\
\midrule
     284 & 287.22968, $-$24.25000 & 48 \\ 
     331  & 262.55696, $-$17.05000& 50\\
     437 &291.41827, $-$2.65000 & 42\\
     553 & 23.12473, 18.95000 & 36\\
     635 & 263.32629, 26.15000& 46\\
     649 & 8.35811, 33.35000& 46\\
     683 &274.70950, 33.35000 & 47\\
     686 &298.26936, 33.35000& 49\\
     687 &306.08789, 33.35000 & 42\\
     737 & 12.25000, 47.75000& 45\\
     757 & 204.72454 , 47.75000& 45\\
     758 &214.49923, 47.75000 & 33\\
     765 & 282.92208, 47.75000& 48\\
     768 &312.24615, 47.75000 & 43\\
     778 &60.11656, 54.95000 & 40\\
     794 & 235.60357 , 54.95000& 39\\
     797 &268.89602, 54.95000 & 47\\
     800 &302.18847, 54.95000 & 47\\
     813 & 100.00000, 62.15000& 38\\
     820 &193.33333, 62.15000 & 45\\
     830 &326.66667, 62.15000 & 38\\
     831 &340.00000, 62.15000 & 40\\
     837 & 77.14286, 69.35000& 41\\
     843 &180.00000, 69.35000 & 37\\
     848 &265.71429, 69.35000 & 48\\
     856 & 60.00000, 76.55000& 47\\
\bottomrule
    \end{tabular}
}
    \caption{Summary of selected fields, including their central coordinates and the number of artefacts (also nominals) from each field included in the presented datasets.}
    \label{tab:fields}
\end{table}

\begin{figure*}
    \centering
    \includegraphics[width=0.67\linewidth]{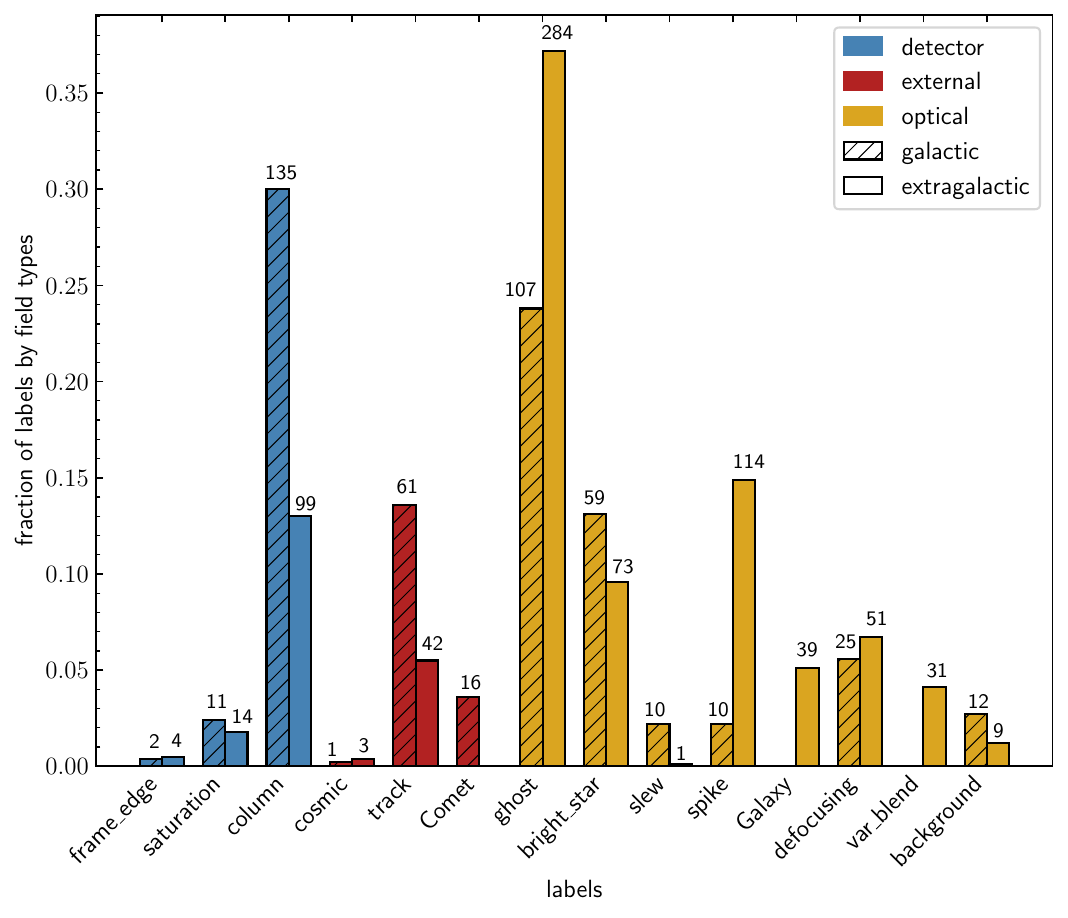}
   \caption{Bar plot showing the distribution of artefact labels in the dataset, separated into galactic and extragalactic fields. The x-axis lists the different artefact subtypes, while the y-axis represents their fraction within each group. The numbers above the bars indicate the total counts of each artefact type. The colour scheme follows the categorization of artefacts into detector-related (blue), external (red), and optical system (yellow) artefacts, as described in Section~\ref{sec:arts} and illustrated in Figure~\ref{fig:schema}.}
    \label{fig:types}
\end{figure*}

In total, we labelled 1127 objects as artefacts with 1213 tags. The difference between the number of objects and the number of tags arises because, in some cases, a single object required up to three labels due to multiple effects occurring simultaneously.
According to the classification defined in Section~\ref{sec:arts}, the artefact dataset includes 265 detector artefact labels, 825 optical artefact labels, and 123 external artefact labels. Among the detector artefacts, the most common subtype is \texttt{column} (234), followed by \texttt{saturation} (25), while \texttt{frame\_edge} (6) is the least common. The most frequent optical artefact label is \texttt{ghost} (391), making it the most common artefact overall. Among external artefacts, \texttt{track} (103) is the most frequent, while \texttt{cosmic} (4) is the least common, making it the rarest artefact label in the dataset. The detailed distribution of artefact subtypes is shown in Figure~\ref{fig:types}.

The final artefact dataset consists of 1127 images in FITS format, available in two sizes: $28 \times 28$ and $63 \times 63$ pixels. Table \ref{tab:tags} lists the additional header entries included in the artefact FITS files, beyond those present in the original ZTF headers. The nominal dataset follows the same format, except for the absence of keyword entries. Both datasets, in both image sizes, are publicly available at Zenodo (\url{https://zenodo.org/records/15076426}).

\begin{table}
\setlength{\tabcolsep}{4pt}
\centering
\scalebox{0.85}{
\begin{tabular}{l|l}
\midrule
Keyword  & Description \\
\midrule
\texttt{OID}& ZTF object ID \\
\texttt{OIDRA} & Object Right Ascension (deg)\\
\texttt{OIDDEC} & Object Declination (deg)\\
\texttt{TAGS} & Labels for the central object according to the schema in Section \ref{sec:arts}.\\
\texttt{URL} & Download URL for the FITS image\\
\\
\bottomrule
\end{tabular}
}
\caption{Table showing the additional entries that have been added to the ZTF FITS headers and their respective descriptions.}
\label{tab:tags}
\end{table}

\section{Discussion} 
\label{sec:disc}

\subsection{Distribution of artefacts by field}

To investigate potential spatial dependencies in artefact distribution, we examined their occurrence in relation to field positions. We classified fields as `galactic' (between $-15$ deg and $+15$ deg in galactic latitude) or `extragalactic', and calculated the fraction of each artefact type in both groups (see Figure \ref{fig:types}). 

Some artefact types, such as \texttt{frame\_edge}, \texttt{column}, and \texttt{ghost}, exhibit differences between galactic and extragalactic fields, but these variations are likely due to chance, as such artefacts originate from the CCD or system-related effects and do not depend on sky position. In contrast, bright-star-induced artefacts are expected to be more frequent in galactic fields, where the stellar density is higher. Indeed, this trend is evident for \texttt{bright\_star} and \texttt{saturation}. However, interestingly, \texttt{spike} and \texttt{var\_blend}, which are also linked to bright stars, are more common in extragalactic fields. The \texttt{Galaxy} tag, as expected, is found only in extragalactic fields, since background galaxies are largely obscured in the Galactic plane.

While these trends suggest potential spatial dependencies, the process of active anomaly detection introduces an additional layer of complexity. A key characteristic of active learning, and particularly of our active AD approach, is that each new decision made by the algorithm is influenced by prior interactions with an expert. This property may introduce a selection bias, affecting which artefacts are detected in certain fields and potentially masking true environmental or spatial dependencies. For example, if an expert labels an outlier as \texttt{ghost}, the algorithm is more likely to prioritize showing similar outliers in subsequent iterations. We observe this effect in some fields. For instance, in our dataset, every image in field 284 contains a ghost artefact. However, it is unclear whether this is due to selection bias or because ghosts are genuinely the most common artefact in that field. In contrast, field 553 exhibits a diverse range of artefacts.

\subsection{Photometry of bright stars from artefact light curves}
\label{mira_echoes}

\begin{figure*}
    \centering
    \includegraphics[width=\textwidth]{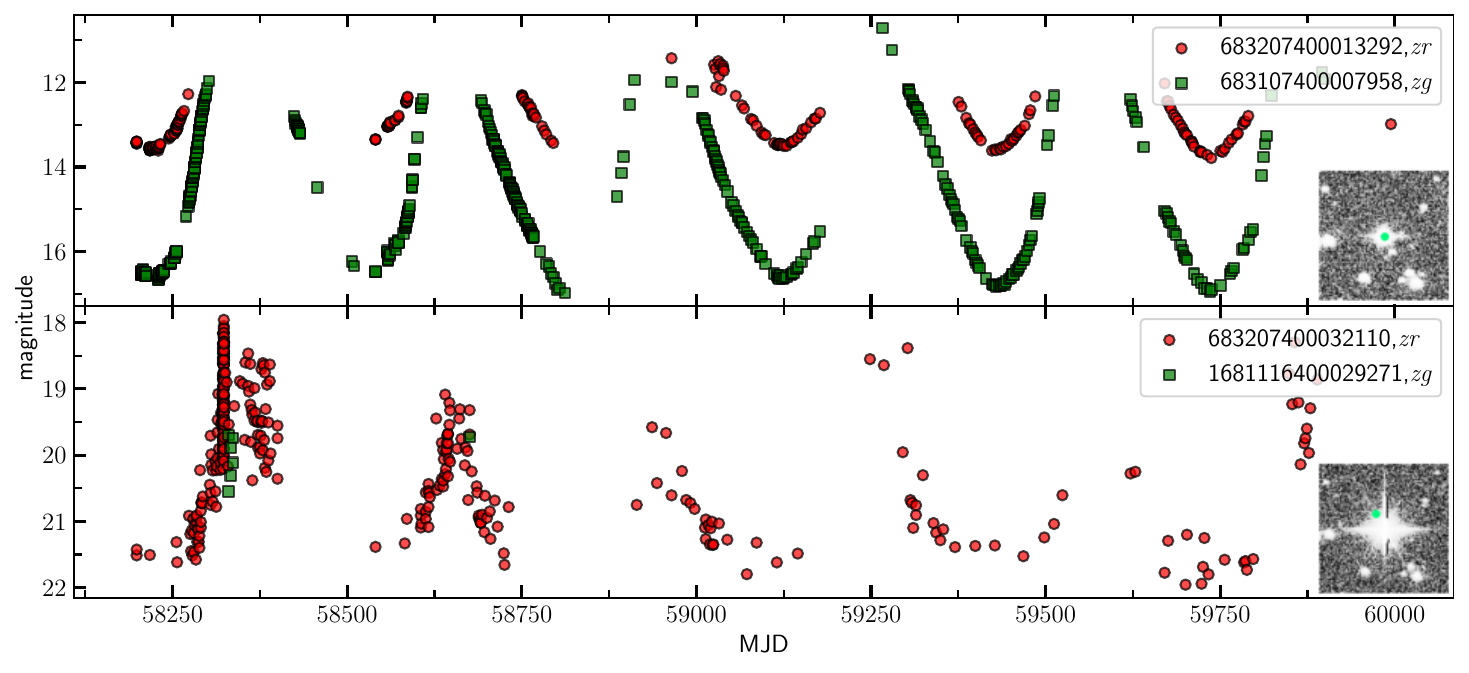}
    \caption{Illustration of the \texttt{var\_blend} artefact label. The top panel shows the \textit{zr} \& \textit{zg}-band light curves of a Mira variable (ZTF OID \texttt{683207400013292}, also known as TV Her), with an inset showing its ZTF FITS image. The bottom panel shows the light curves of an object tagged as \texttt{var\_blend} (ZTF OID \texttt{683207400032110}), which represent scattered echoes of the same Mira variable in a different exposure. The inset images are 100 $\times$ 100 pixels in size, with a target marked by a green dot. Astrometric directions follow the standard convention (North up, East left).}
    \label{fig:miraecho}
\end{figure*}

Some artefact light curves closely resemble those of real variable stars or transients, making it difficult to distinguish them from genuine astrophysical variability based on photometry alone. A key example is Mira-variable echoes, where light from a bright variable star contaminates a nearby source, creating the illusion of its variability. Figure~\ref{fig:miraecho} illustrates this effect. The top panel shows the \textit{zg}- and \textit{zr}-band light curves of the Mira variable TV Her, with its FITS image inset. The bottom panel presents a similar light curve and image for an object we labelled as \texttt{var\_blend}. While the artefact light curve appears periodic, only the image reveals the true cause: contamination of nearby non-variable objects by light from a bright variable star (TV Her in this case). A similar issue occurs with \texttt{track} artefacts, where a single anomalous point in an otherwise normal light curve might resemble a transient event, such as a stellar flare, until the image reveals it as a satellite track. This highlights the importance of combining light curve analysis with image inspection to correctly identify artefacts.

Moreover, \texttt{var\_blend} artefacts offer an interesting scientific application. If a bright variable source is too saturated\footnote{According to the \citealt{2009arXiv0912.0201L} estimates, saturation for a 15-second exposure in 0.7 arcsec seeing occurs at magnitudes of $u$, $g$, $r$, $i$, $z$, $y$ $=$ 14.7$^m$, 15.7$^m$, 15.8$^m$, 15.8$^m$, 15.3$^m$, and 13.9$^m$, respectively. Future large-scale surveys such as LSST are expected to contain a significant number of such cases.} to be properly measured and is therefore absent from the survey data releases, its variability can still be inferred from the photometry of nearby contaminated sources.
Consider the example of the Mira variable TV Her and its echos shown in Figure~\ref{fig:miraecho}. Using the Lomb-Scargle periodogram \citep{lomb1976,scargle1982}, we estimated the period of the \texttt{var\_blend} artefact (ZTF OID \texttt{683207400032110}) to be 303.6 days, which is in remarkable agreement with the true period of TV Her (304.3 days\footnote{\url{https://www.aavso.org/vsx/index.php?view=detail.top\&oid=14840}}). This demonstrates that even artefact light curves can provide additional valuable astrophysical information about bright sources.

\subsection{Interesting cases}
\subsubsection{Comet 21P/Giacobini-Zinner}

We identified 16 instances of Comet 21P/Giacobini-Zinner within field 778 of our dataset. According to our labelling system, these are classified as external artefacts, as the distortion in the target light curve results from overlap with the comet's light. Interestingly, such detections could serve as an indirect method for discovering new comets.

A video showing the motion of Comet 21P/Giacobini-Zinner from 2018-08-20 to 2018-09-20 can be viewed on the SNAD YouTube channel\footnote{\url{https://youtube.com/shorts/AZtcAL5e5fA?feature=shared}}.

\subsubsection{New eclipsing binary}

\begin{figure*}
    \centering
    \includegraphics[width=0.8\textwidth]{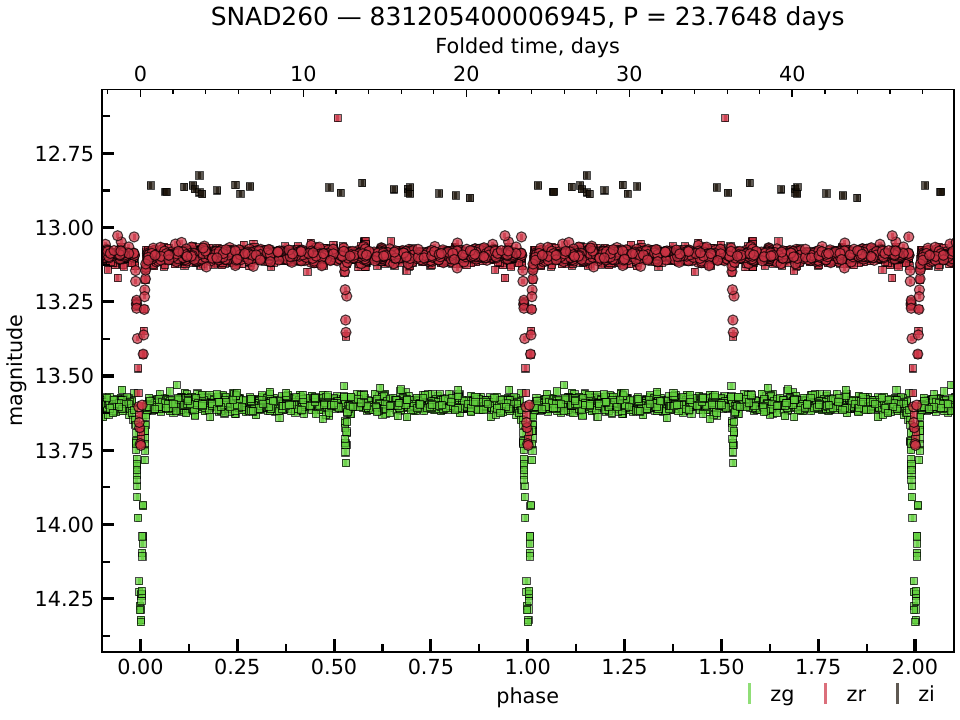}
   \caption{The folded light curves of eclipsing binary SNAD260. ZTF OIDs: \texttt{831205400006945, 831105400004328, 831305400076508, 1837114400010913, 1837214400022074, 832208300003915, 832108300002373}.}
    \label{fig:EB}
\end{figure*}

During this work, we discovered a previously uncatalogued eclipsing binary with ZTF OID \texttt{831205400006945} (Fig.~\ref{fig:EB}). The object has been added to the AAVSO VSX\footnote{\url{https://www.aavso.org/vsx/index.php?view=detail.top&oid=10319899}}, the SNAD catalog\footnote{\url{https://snad.space/catalog/}} and assigned the internal name SNAD260. Maximum and minimum magnitudes are 
13.09$^m$ and 13.73$^m$ in $zr$-band, respectively. Its period, determined using the Lafler \& Kinman method \citep{Lafler_Kinman}, is 23.7648 days.

\section{Summary and conclusions} 
\label{sec:summconc}

In this work, we present a dataset of artefacts detected in the Zwicky Transient Facility DR3 using the \pine \  active anomaly detection algorithm. We selected 26 ZTF fields, applied anomaly detection with a budget of 50 iterations per field, and labelled the detected artefacts following the SNAD classification schema. The final dataset consists of 1127 artefact images with a total of 1213 labels, as some objects exhibit multiple artefact types. Additionally, we provide a nominal dataset of 1127 nominal objects from the same fields for comparison. Both datasets are available in FITS format in two image sizes,  $28 \times 28$ and $63 \times 63$ pixels.

The most common artefact type in our dataset is \texttt{ghost}, followed by \texttt{column} and \texttt{bright\_star}, while the least frequent are \texttt{cosmic} and \texttt{frame\_edge}. We investigated the spatial distribution of artefacts and found no clear correlation between artefact type and sky position. The classification of artefacts by active learning introduces an additional selection bias, as the algorithm prioritizes anomalies similar to those previously labelled by an expert. 

In addition to the creation of the dataset of artefacts and nominal objects, we demonstrated a practical scientific application of artefact light curves. In particular, we showed that the variability period of a bright, saturated source can be recovered from the light curves of nearby contaminated objects. As an example, we measured the period of an artefact light curve contaminated by the light of a nearby Mira variable and found remarkable agreement with the true period of the Mira variable. We also reported the identification of 16 instances of Comet 21P/Giacobini-Zinner. This suggests that similar detections could provide an indirect method for identifying new comets. Finally, we discovered a previously uncatalogued eclipsing binary (ZTF OID \texttt{831205400006945}), which has been added to the SNAD catalog under the designation SNAD260.

All astronomical surveys are affected by artefacts, making their automated identification and removal essential for maintaining high-quality catalogs. Machine learning approaches, particularly supervised learning methods, rely on labelled training data, which is often a limiting factor. While some studies use synthetic artefacts for training and citizen science initiatives such as Zooniverse, expert-labelled datasets remain a valuable resource for training and validating machine learning models. The datasets provided in this work can be used for real-bogus classification, anomaly detection, and survey catalog cleaning. Moreover, \pine \  active anomaly detection approach can be further adapted for targeted artefact identification in large surveys. Future applications may include LSST-era surveys, where the increasing volume of astronomical data will make automated artefact identification even more critical.

\section*{Acknowledgements}

The study was conducted under the state assignment of Lomonosov Moscow State University.
Support was provided by Schmidt Sciences, LLC. for K. Malanchev. The work of V.~Krushinsky was supported by a project of youth scientific laboratory, topic FEUZ-2025-0003.

\bibliographystyle{elsarticle-harv} 
\bibliography{example}

\end{document}